# Modelling trade offs between public and private conservation policies

Ascelin Gordon[a,*] William T. Langford[a], Matt D. White[b], James A. Todd[c], Lucy Bastin[d]

[a]School of Global Studies, Social Science & Planning, RMIT University, GPO Box 2476, Melbourne 3001 Australia. *bill.langford@rmit.edu.au*.

[b]The Arthur Rylah Institute for Environmental Research, Department of Sustainability and Environment, PO Box 137, Heidelberg 3084, Australia. *matt.white@dse.vic.gov.au*.

[c]Biodiversity and Ecosystem Services Division, Department of Sustainability and Environment, 2/8 Nicholson St East Melbourne 3002, Australia. *james.todd@dse.vic.gov.au*.

[d]School of Engineering and Applied Science, Aston University, Birmingham B47ET, UK. *l.bastin@aston.ac.uk*.

[*]Corresponding author: Tel.:+613 99259930 Fax:+613 99253088
*E-mail address: ascelin.gordon@rmit.edu.au*

**Abstract**

   To reduce global biodiversity loss, there is an urgent need to determine the most efficient allocation of conservation resources. Recently, there has been a growing trend for many governments to supplement public ownership and management of reserves with incentive programs for conservation on private land. At the same time, policies to promote conservation on private land are rarely evaluated in terms of their ecological consequences. This raises important questions, such as the extent to which private land conservation can improve conservation outcomes, and how it should be mixed with more traditional public land conservation. We address these questions, using a general framework for modelling environmental policies and a case study examining the conservation of endangered native grasslands to the west of Melbourne, Australia. Specifically, we examine three policies that involve: i) spending all resources on creating public conservation areas; ii) spending all resources on an ongoing incentive program where private landholders are paid to manage vegetation on their property with 5-year contracts; and iii) splitting resources between these two approaches. The performance of each strategy is quantified with a vegetation condition change model that predicts future changes in grassland quality. Of the policies tested, no one policy was always best and policy performance depended on the objectives of those enacting the policy. This work demonstrates a general method for evaluating environmental policies and highlights the utility of a model which combines ecological and socioeconomic processes.

**1 Introduction**

Globally, there has been an increasing loss of biodiversity and habitat due predominantly to anthropogenic land-use change (Millennium Ecosystem Assessment, 2005). To ensure a sustainable future, there is an urgent need to determine the most efficient allocation of resources for conserving biodiversity (Wu and Boggess, 1999). This is a complex task, as determining the most effective conservation actions or policies for a given region involves balancing ecological, financial, and social constraints.

Conservation actions that target biodiversity on private land can act as vital supplements to conservation that targets public land, and private land conservation is increasingly recognized as being strategically important (Figgis, 2004; Knight, 1999; Newburn et al., 2005). Both in Australia and internationally, there are areas where a significant proportion of native and/or ecologically important vegetation occurs on private land (Drechsler et al., 2007; Figgis, 2004; Knight, 1999). This has led government agencies around the world to explore a range of mechanisms for promoting conservation on private land, such as fixed price grants, tax incentives or voluntary schemes (Doremus, 2003; European Commission, 2005; Main et al., 1999; USDA, 2010). More recently in Australia, conservation contracts have been auctioned within the *BushTender* scheme run by the Victorian Government's Department of Sustainability and Environment (DSE) (Stoneham et al., 2003; DSE, 2008a). Delivering such a scheme involves consulting with the landholders in a given area and having them offer a price (bid) for undertaking conservation actions on their property over a fixed time. Participation is voluntary and the agency is then able to choose which bids to accept or reject. Schemes such as these have the potential to deliver conservation outcomes in a more cost effective manner than other fixed-price methods of private land conservation (Stoneham et al., 2003).

Although policies to promote conservation on private land have been implemented in numerous countries, they are rarely evaluated in terms of their ecological consequences (although there are exceptions, e.g. Natural England (2009)). This means that policy choices are often made in a non-transparent manner based on some mix of conjecture and expert opinion, rather than on evidence. Here, we present an illustrative example and a method for assessing the consequences of conservation policies that incorporates both ecological and socioeconomic processes, and allows policy outcomes to be evaluated based on clear and transparent assumptions.

Specifically, we focus on an example comparing the outcomes of private land conservation schemes that involve short-term conservation contracts (such as *BushTender*) to the outcomes of more traditional methods such as purchasing land to establish public conservation areas. Although the analysis presented here models one specific type of private land conservation in an Australian context, the modelling approach used is general. It could be adapted to other contexts that involve the allocation of conservation resources among a group of stakeholders, with ensuing biodiversity gains or losses. A re-parameterisation of the model components with empirical data and/or expert opinion would allow its application to other locations (e.g., European agri-environmental schemes or US Landowner Incentive Programs) or to additional private conservation alternatives such as easements or covenants.

We demonstrate our approach with a case study examining the conservation of the Western (Basalt) Plains Natural Temperate Grassland (henceforth called *native grassland*) to the west of Melbourne, Australia. This native grassland is one of Australia's most endangered vegetation types with over 99.5% of its original distribution having been lost or substantially altered (Williams et al., 2005). It supports a diversity of animal species (DSE, 2004), including eight nationally threatened species. The remaining remnants predominantly occur on private land (DNRE, 1997) and continue to face a number of threats including inappropriate grazing or fire regimes, weed invasion, inappropriate herbicide use and the application of fertilizers (DSE, 2004; Carter et al., 2003). Thus it is thought that most native grassland on private land will have a deterministic decline in condition if left unmanaged (Williams et al., 2006). Further background on the ecology and history of these native grasslands are provided in the Supplementary information.

In our case study, we explore the situation where a fixed budget is available every 5 years that can either be spent completely on public conservation (involving land purchase and managing existing reserves), private land conservation (*BushTender* type conservation contracts) or an equal mix of both. We examine how each of these strategies performs, relative to a scenario with no intervention. We also examine how the policies perform in the presence of development, which is assumed to destroy native grassland. The performance of each strategy is quantified with a grassland condition model that predicts how the vegetation quality of grassland changes with time depending on its current condition and the management status of the vegetation.

<< Fig. 1 >>

## 2 Methods

We used a previously-developed general modelling framework to simulate the ecological and socioeconomic processes involved with each conservation policy (Langford et al. 2009). The framework consists of the following 6 steps: 1) define study area, 2) define planning units and habitat information, 3) define costs and budgets, 4) undertake conservation actions, 5) model system dynamics, and 6) measure consequences. Below we describe how each step relates to the present analysis. More details on the modelling framework can be found in Langford et al. (2009).

*2.1 Define study area*

The study area is located near Melbourne, Australia (Fig. 1). It borders the western growth boundary of Melbourne and is predominantly agricultural land, but also contains regions of urban and industrial land-use. The study area was chosen because it is an area where the Victorian Government plans to purchase land for public conservation to offset the loss of native grassland from the proposed expansion of Melbourne (DPCD, 2009). Together with DSE, we were interested in developing models to understand how private land conservation policies might supplement public land conservation in this area.

*2.2 Define planning units and habitat information*

Habitat information was obtained from a map depicting the condition of the Western (Basalt) Plains Natural Temperate Grassland community (Fig. 1). The map was derived from a modelled vegetation condition layer for the state of Victoria (GIS layer NV2005_QUAL1; DSE, 2009a) and site data collected during 2008-2009 by the Victorian Growth Areas Authority and DSE. The modelled vegetation condition layer was generated from site measurements of vegetation composition and structure in conjunction with coincident environmental data to build an extensive spatially explicit model of vegetation condition across the state of Victoria (Kocev et al., 2009; Newell et al., 2006). Each cell represents the condition of grassland as measured by the *habitat hectares* vegetation assessment method (Parkes et al., 2003). This score represents vegetation condition relative to a mature and undisturbed benchmark of the same vegetation type. *Habitat hectares* scores range between 0 and 100, and include a site condition component (75%) and a landscape context component (25%). For this analysis, only the site condition component of the score was used and the grassland condition map was divided by 100 producing condition scores ranging between 0.0 and 0.75 (Fig. 1).

The planning units define the discrete elements upon which policies can act. They consist of properties derived from Victoria's Cadastral Area Boundary layer (DSE, 2009a). Properties with an area less than 5 hectares were excluded from the study, as these properties are primarily residential and contain small amounts of lower quality grassland (parcels greater than 5 hectares contain 93% of the native grassland in the study area, and 95% of the higher quality grassland; see Supplementary information for more details). This resulted in 3302 properties with a median size of 13.3 hectares, a mean size of 33.7 hectares and a standard deviation of 102 hectares.

All data was converted to raster format with a 50 × 50 $m^2$ pixel resolution. This resolution was chosen so that grassland condition variation could be clearly resolved in the smallest land parcels.

*2.3 Define costs and budgets*

The purchase price of a parcel was generated by first obtaining a price per hectare, and then using this value to calculate the full parcel price. The price per hectare was determined by sampling from a lognormal distribution based on real sale price data from the study area and surrounds. As larger parcels tend to have a lower price per hectare, we split the real price data into two samples, depending on whether the parcel area was less than or greater than 150 hectares, and then fitted a lognormal distribution to each sample. The parcel price per hectare in the model was then determined by sampling from the appropriate lognormal distribution for each parcel. This allowed us to achieve a realistic variation in the sale prices of the parcels being modelled, as well as capturing some of the correlation between sale price per hectare and parcel area (see Supplementary Information for further details).

Management costs were based on the known costs of managing similar grassland in other areas of the state and were given a fixed value per hectare. In a *BushTender* trial in the Victorian Riverina Bioregion, management costs paid to land holders with properties containing Northern Plains Grassland were similar to the cost of ongoing management of existing conservation reserves in the area (DSE, unpublished data). Using this information as a guide to realistic management cost ranges, we explored scenarios with two different management costs for a 5-year time period for both public and private conservation: *low* ($100 per hectare) and *high* ($300 per hectare).

At each time-step a fixed budget was allocated to the policy being undertaken. Three budgets were used to investigate the impact of budget size on policy ranking: *small* ($1 M), *medium* ($1.5 M) and *large* ($2 M). These values were used to reflect plausible budgets for both public and private land conservation. Previous *BushTender* trials have had a budget of over $0.9 M (DSE, 2009b) and a $2 M budget per 5 years is realistic for public conservation on the western edge of Melbourne, considering that $50 M was spent on land acquisition for extending the parkland network throughout all of metropolitan Melbourne between 2005 - 2008 (DSE, 2008b).

*2.4 Undertake conservation actions*

Simulations were run for 80 years, with 16 time-steps, each of five years duration. These settings were chosen to allow the long-term consequences of policies to play out over time, with a temporal resolution set to the length of a private conservation contract. At each time-step, a policy was followed until the budget was exhausted. Each policy was run 5 times with a different random seed, to explore the extent to which model stochasticity caused results to vary.

We examined four policies for undertaking conservation actions, covering the extremes and midpoint of how budgets could be split between policies:

i) *Conservation on private land only* - Undertake conservation by paying a random subset of landholders for a 5-year (single time-step) contract to manage the native grassland on their land. Once the contract expires, the landholder is eligible to enter into a new contract. This policy models a voluntary program that is dependant on willing landholders choosing to participate. The random selection of landholders simulates the likelihood that participation will not be related to the quality or quantity of grassland on their parcel (see section 4.3 for further discussion of this point).

ii) *Conservation of public land only* - First pay to manage existing public reserves for the next time-step (if any), and then spend the remaining budget on purchasing and managing new land parcels. With this strategy, a point may be reached where the entire budget will be spent managing existing reserves. As an agency would attempt to be strategic in purchasing land for conservation, we used the following greedy heuristic for ranking and selecting available parcels. First, land parcels with a mean condition score greater than the grassland condition threshold (see below) are ranked by their benefit/cost score, calculated by summing the condition score of all grassland pixels in the parcel and dividing it by the purchase cost. Parcels are then purchased and managed in order until the budget is exhausted, or until no parcels above the threshold remain. If budget remains, the same procedure is applied to parcels below the condition threshold.

   iii) *Mixed strategy* - The budget is split equally between undertaking conservation on private and public land. For each portion of the budget, the relevant purchase/management strategies above are followed.

   iv) *Do nothing* - There are no conservation interventions in the system.

*2.5 Model system dynamics*

To estimate the effects of conservation actions, we used a model describing how the condition of native grassland changes depending on its management status. All grassland is assumed to be in one of three categories: *managed*, *unmanaged*, or *developed*. Land is considered to be *managed* when it is within a public reserve or private land incentive program. *Unmanaged* land is grassland on private land subject to entitled uses (such as grazing) and uncontrolled threats (such as perennial weed invasion). *Developed* land is assumed to no longer support viable native grassland, due to urban development.

Future grassland condition was calculated at the per-pixel level. At each time-step the condition of grassland was updated using the curves shown in Fig. 2. These curves reflect behaviour specific to the grasslands in this study and were parameterised using the expert opinion of DSE ecologists. Unmanaged grasslands were assumed to degrade over time, while managed grasslands were assumed to improve in quality. However, once a patch of grassland falls below a certain condition threshold, it is difficult to fully restore it, so there are two different condition curves for managed land. The most likely value of the threshold was estimated at 0.35, based on DSE expert opinion. The two solid curves in Fig. 2 show how the condition of a pixel of managed grassland will change over time depending on its initial condition: if it starts with a condition score above/below 0.35, it will asymptote towards a value of 0.75/0.35, respectively.

To model stochasticity of the condition change process, the condition score of each pixel was perturbed by adding a small random value sampled from a normal distribution with a mean of zero and a standard deviation of 0.02 (based on DSE expert opinion). A small proportion of pixels were allowed to cross the threshold due to these random fluctuations and embark on the higher recoverability curve. The probability of this occurring was 0.0005, based on the expert opinion of DSE ecologists.

As we were interested in investigating how policy performance varied in the presence of adversity, scenarios were run with development that is assumed to destroy vegetation. In these scenarios, 207 randomly selected parcels were developed at each time-step. We chose this development rate so that all unprotected parcels would be developed during the 80-year period of the simulation. Parcels were selected randomly because there is currently little information on likely development patterns over the next 80 years and much of the area has homogenous (rural) land-use zonings.

<< Fig. 2 >>

*2.6 Measure consequences*

This final step quantifies the consequences of the modelled actions by summarising the output of the grassland condition change model. Two metrics were used to obtain aggregate scores of grassland condition at each time-step: (i) the *total summed condition* (TSC), calculated by summing the condition value of all pixels of grassland in the study area; (ii) the summed *condition above threshold* (CAT), obtained by summing the values of each pixel of grassland in the study area with a condition score higher than the 0.35 threshold (see section 2.5). These metrics were chosen to give insight into how given conservation policies affect the extent and condition of native grassland in the study area. The CAT metric provides an indication of the extent and quality of grassland that has the potential to reach a high condition score.

**3 Results**

The four policies under investigation were: (i) spend the entire budget on public conservation (ii) spend the entire budget on private conservation, (iii) split the budget equally between the two in a mixed strategy, or (iv) do nothing.

Figs. 3 – 5 show how aggregate measures of native grassland extent and condition vary with time using the *small* and *large* budgets. In Figs. 3 – 5 the left column shows the *total summed condition* (TSC) of all grassland in the study area on the *y*-axis while the right column shows the summed *condition above* the 0.35 *threshold* (CAT). Results are shown for *low* management cost in Figs. 3 and 4 and for *high* management cost in Fig. 5. The median condition value of each policy is shown as a solid or dashed line, and the grey band depicts the range of trajectories for that policy over the five realisations of the simulation. In some situations there is considerable overlap between bands.

The results with no development are shown in Fig. 3. When evaluating the policies using TSC and the small budget, private conservation performs best, followed closely (in the long term) by the mixed strategy and then public conservation (Fig. 3a). The difference in performance between the three strategies is exaggerated when the budget is increased (Fig. 3c). However, if the higher quality grassland is used to evaluate the policies (using the CAT metric), there is a crossover point where public conservation overtakes the private and mixed strategies (Fig. 3b, 3d). For the small budget, this crossover point occurs at approximately 32 years. For the large budget, the private conservation policy still has a greater score at the end of the 80-year simulation, but eventually falls below public conservation if the simulation time is increased. In both CAT-evaluated plots the mixed policy has considerable overlap with the public conservation policy, with a median condition score only slightly below public conservation at the end of the simulation. In all cases the mixed budget also yields a higher condition score for the first 25 years compared to public conservation.

<< Fig. 3 >>

Policy performance was also explored in the presence of development (Fig. 4), to gauge the robustness of policies to this particular type of adversity. The additional stochasticity from the development model increases the performance variability of all policies under all budgets, denoted by the thickening of all the grey bands in Fig. 4. Evaluating the policies with the CAT metric (Fig. 4b, 4d) shows only a small change compared to the case without background development (Fig 3b, 3d). The main difference is in the performance of private conservation with the larger budget (Fig. 4c and d), where performance drops off more rapidly towards the end of the simulation, compared to the case without development. When TSC is used to evaluate the policies, development has a larger impact (Fig. 4a, 4c) resulting in highly degraded performance for all policies. The policies tend to have a greater overlap in performance, especially where the budget is small (Fig. 4a). A change in ranking also occurs relative to the case without development, where public conservation overtakes both the private and mixed policies after approximately 60 years (Fig. 4a, 4c).

<< Fig. 4 >>

We also explored the consequences of increasing the management cost per 5-year period to the higher value of $300 per hectare (Fig. 5). This threefold increase in management outlay has a significant influence, with all policies suffering differing reductions in performance. Public conservation, where land purchase dominates the budget, only suffers a small reduction in performance. In comparison, private conservation suffers the largest reduction as the whole budget is spent on land management, while the mixed policy suffers an intermediate reduction. With increased management cost, both the mixed and private conservation policies end up falling below the public policy for all budgets and evaluation criteria (Fig. 5).

<< Fig. 5 >>

The results in Figs. 3 – 5 were also run with the medium budget of $1.5 M and these results are shown in the Supplementary information (Figs. S3 and S4), along with the *small* and *large* budget results for reference. The medium budget results are consistent with a transition between the *small* and *large* budgets.

**4 Discussion**

Deciding how to allocate limited conservation budgets across a range of actions can be challenging. Predicting the consequences of any given allocation strategy can be equally difficult. The study discussed here demonstrates a process for exploring the consequences of allocating a fixed budget between different policies involving conservation on both public and private land. This process combines a method for implementing a complex sequence of actions with a system model that simulates the response of the landscape to those actions.

In presenting this case study we aimed to demonstrate the feasibility and utility of a complex sequential model for examining policy performance and robustness to a range of scenarios and future adversities. Each sub-component of our model has the potential for increased realism and complexity, or could be altered to model other policies, locations or vegetation types. The design of the modelling framework we used is modular for this very reason (Langford et al., 2009). Later in the discussion we consider some further assumptions in our model, and ways of incorporating added complexity and realism. One of our goals in presenting this study is to demonstrate the feasibility and usefulness of one such method, with the hope of encouraging the increased use of evaluation and evidence in the development of conservation policy.

*4.1 Impact of policy objectives*

Comparing policies inherently depends on the objectives of those who are enacting policy. The results in Figs. 3 – 5 show that there are tradeoffs between policies depending on budget, management costs, time horizon and how quality and future persistence of grassland is valued, relative to the gross quantity of grassland.

The private conservation strategy resulted in a larger area of grassland being managed at any given time because there were no land acquisition costs. However, the area being managed changes as contracts expire and new ones are initiated, so a given parcel may potentially move in and out of management several times throughout the simulation. The amount of time that a parcel is left unmanaged, with a resulting decrease in grassland condition, depends on the proportion of the landscape being managed at any given time-step, which in turn depends on the budget and available area of grassland. If only a small proportion of the landscape can be managed at any time, parcels are likely to spend little time in management and to fall below the condition threshold (Fig. 3b, line 2). This reduces the effectiveness of future management, since once grassland falls below the condition threshold, it is likely to remain in poor condition regardless of the frequency or expense of subsequent management.

Public conservation takes the opposite approach where land is purchased and retained and these parcels stay managed throughout the simulation. This strategy provides a smaller, but steadily increasing amount of grassland that is, on average, of higher quality.

The information in Figs. 3 – 4 can be used to evaluate a given scenario's efficacy in providing habitat for species with a range of grassland habitat requirements. If the sole objective was to maximise the *amount* of grassland over the next 20 years, then in some situations private conservation outperformed public for both large and small budgets (Fig. 3a, 3c). If the objective was to maximise the amount of high quality grassland in 80 years time, then there were situations were public conservation was the preferable method under *small* and *medium* budgets (Figs. 3b, S3) and also the preferable method under the *large* budget if a longer time horizon was considered (Fig. 3d). In ecological terms, if a species of concern could utilise grassland in a range of degraded conditions, then the private conservation strategy could provide more habitat for that species provided development was minimal and management costs were not large. However, if a species required grassland in high condition then public conservation provided a greater amount of secure habitat. For some objectives, the mixed strategy could be preferable, since at the end of the simulation the mixed strategy often had only a slightly lower score than public conservation but had a significantly higher score earlier in the simulation (e.g. Figs. 3d, 4d).

With high enough rates of development, it would always be preferable to have at least some component of public conservation, due to its protective effect. Where large amounts of habitat are vulnerable to loss due to the limited protection of private conservation contracts, development pressure causes the private conservation policy to end up with the same low summed condition score as the *Do nothing* scenario (Fig. 4).

*4.2 Purchase and management costs*

Purchase and management costs were an important factor in driving policy performance. In real peri-urban situations, parcel purchase costs per hectare can vary considerably due to many context-specific factors. We incorporated this variation into the model by stochastically determining parcel purchase prices in a way that is consistent with the distribution of real parcel sales data from the study area (see Supplementary information). A significant portion of the variation in policy performance between model realisations was due to this stochastic assignment of costs. This variation is useful for assessing how a policy's performance might change if applied to other urban-fringe locations with similar vegetation, where the purchase costs had comparable distributions.

Varying the management costs had a significant impact on all policies. High management costs (Fig. 5) changed the ranking of policies when using TSC to evaluate performance (Fig. 5a, 5c). This demonstrates that model outcomes can be sensitive to management costs and that a case study using vegetation with different management costs could significantly alter predicted and real policy performance.

Finally, management costs were assumed to be the same for both public and private land. While this may be approximately true, several factors could cause management costs to vary. On public land, management costs will be sensitive to reserve aggregation due to the increased resources required for a single agency to manage many small, dispersed parcels. Similarly, the management cost for private conservation contracts offered by landholders via auction might vary due to landholder competition and differing implementation and opportunity costs. These factors could be incorporated to investigate their effect on policy performance, though it would incur a considerable increase in model complexity.

*4.3 Assumptions and generalization*

In this study, parcels chosen for private conservation were selected randomly, which may be realistic if an agency's primary goal is to educate and include as many landholders as possible. In reality, while the landholders interested in private conservation could be (effectively) random, the subset of landholders selected for a given program may not be. Victoria's *BushTender* program (Stoneham et al., 2003; DSE, 2008a) uses a strategic cost-benefit approach in selecting a subset of landholders to be awarded conservation contracts. This strategy attempts to maximise participation (and total area) in the program, while also prioritising parcels containing a number of attributes including higher quality vegetation and security, the presence of threatened species and proximity to existing public reserves and private land conservation. This approach, where anyone in the study area has an opportunity to participate and a final selection is made with predefined criteria, deals with issues such as equity and incomplete information, where the agency may not initially know where the greatest benefits lie. Incorporating these refinements would add considerable complexity to the parcel selection procedure for private conservation and could lead to improved private performance, but we have not tested this here.

It is also likely that there would be some proportion of landholders who would never be interested in participating in private conservation contracts. Depending on the conservation value of their land, this could potentially degrade the performance of private conservation. Analogously, there may be cases where landholders are unwilling to sell land to an agency for public conservation. If governments are unwilling to risk the social impacts of compulsory acquisition mechanisms, this could reduce the ability of an agency to strategically purchase land for public conservation and thereby reduce the performance of public conservation.

This does suggest that repeatedly targeting the same landholders would improve the performance of private conservation, allowing it to mimic the acquisition and condition maintenance benefits of the public approach. The question then becomes, how much payment is required to make this worthwhile for the landholder? We have not investigated this question here, as it may also be vulnerable to the temporal uncertainties discussed below in section 4.4. Victoria's *BushTender* program deals with this issue by offering landholders the opportunity to choose a "permanent on-title agreement" as part of their contract. Data from *BushTender* trials indicates that approximately 20% of offered contracts are permanent (DSE, 2006). Incorporating this fact into our model would improve the performance of private conservation, provided that a permanent contract is assumed to entail permanent management.

The management costs of fragmented habitats have been touched on above, and it is commonly assumed that only a public model of conservation allows the strategic management of connected areas. However, given the desirability of habitat connectivity for many threatened species, it is possible that aggregation in private land conservation could also be encouraged by systematically-designed payment schemes which favour particular spatial structures (e.g., Hartig et al. (2009)). Such an approach could theoretically increase biodiversity value for the same expenditure on management contracts. To incorporate measures such as these in our simulations would probably entail an extension to agent-based models, with the associated problems of validation and model tractability. However, it is a possible direction for future research.

Finally, the grassland condition change model is an important driver of model outcomes. In these experiments we have used three different curves to represent the change in condition over time derived from expert opinion. The slope of each curve determines how rapidly condition change occurs as parcels are managed or unmanaged. The condition change interacts with both the length of contract and the frequency of contracts that revisit the same site. The threshold behaviour also affects outcomes because if there was no lower threshold, then every piece of land would be worth investing in, and the low quality parcels would offer the biggest gain per dollar spent. An even bigger change in policy performance might be likely if unmanaged vegetation did not degrade, though this is not realistic for the peri-urban grasslands in this study. When considering other ecosystem types that are less prone to degradation if left unmanaged, protection rather than management may be found to be the primary driver of total landscape condition.

*4.4 Robustness to uncertainties*

The sensitivity of the models and policies to underlying assumptions raises the issue of uncertainties and robustness in general, especially relating to temporal uncertainties. Here we incorporated development as one particular type of adversity and discussed the robustness of policies to this process. Most policies and objectives do not explicitly include a goal of robustness with respect to uncertainty. Yet, some form of adversity and even catastrophe are nearly guaranteed during the course of a policy enacted over many years, as the last 30 years of boom and bust economics demonstrates.

Consider the consequences for policies based entirely on private land management that are limited to parcels where landholders choose to participate. These policies are subject to the political whims of changing governments and as such, provide no guaranteed long-term security for funding contracts. This lack of robustness to uncertainty may counter some of the initial attractiveness of private conservation. However, the alternative policies, based on the consolidation of conservation into a smaller set of public lands, may leave reserves more subject to environmental catastrophes and will still require ongoing government funds for management. These types of arguments could be used for advocating a mixed policy of public and private conservation. Future modelling work will include additional types of adversity to determine whether the best policy under stationary conditions is still the best policy under volatile conditions.

**5 Conclusions**

Many governments are increasingly supplementing public ownership and management of conservation resources with market-based conservation policies enacted on private land. This raises a number of questions such as: Do conservation outcomes improve if conservation is done privately rather than publicly? Is there value in splitting resources between the two approaches instead of committing exclusively to one or the other? If so, how should they be split? We have demonstrated a modelling approach for evaluating conservation policy outcomes and compared the performance of three different policies. No one policy was always best, and performance depended on the scenario being tested and the objectives of those enacting the policy. This highlights the utility of a model that incorporates both the ecological and the socioeconomic process in order to explore the efficacy of conservation policies in specific real-world situations. Importantly, these kinds of models could also help to assess policy robustness with respect to unforeseen adversity and catastrophes in the future.

**Acknowledgments**

We would like to thank Nick Williams and the Victorian Government's DSE for providing data and expert opinion. This research was funded by the Australian Research Council through Linkage Project LP0882780, and the Applied Environmental Decision Analysis research hub (through the Australian Commonwealth Environment Research Facilities programme). We are also grateful to three anonymous referees for comments that improved the manuscript.

**Appendix A. Supplementary information**

Supplementary information associated with this article on grassland ecology, parcel cost and results can be found at the end of this document.

Natural England, 2009. Agri-environment schemes in England 2009. <http://www.naturalengland.org.uk/Images/AE-schemes09_tcm6-14969.pdf accessed 14.7.2010>.

Newburn, D., Reed, S., Berck, P., Merenlender, A., 2005. Economics and land-use change in prioritizing private land conservation. Conservation Biology 19, 1411-420.

Newell, G.R., White, M.D., Griffioen, P., Conroy, M., 2006. Vegetation condition mapping at a landscape-scale across Victoria. Ecological Management & Restoration 7, S65-S68.

Parkes, D., Newell, G., Cheal, D., 2003. Assessing the quality of native vegetation: The 'habitat hectares' approach. Ecological Management & Restoration 4, 29-38.

Stoneham, G., Chaudhri, V., Ha, A., Strappazzon, L., 2003. Auctions for conservation contracts: an empirical examination of Victoria's BushTender trial. The Australian Journal of Agricultural and Resource Economics 47, 477-500.

USDA, 2010. United States Department of Agriculture Conservation Programs (including Landowner Incentive Schemes). <http://www.nrcs.usda.gov/programs accessed 14.7.2010>.

Williams, N.S.G., McDonnell, M.J., Seager, E.J., 2005. Factors influencing the loss of an endangered ecosystem in an urbanising landscape: a case study of native grasslands from Melbourne, Australia. Landscape and Urban Planning 71, 35-49.

Williams, N.S.G., Morgan, J.W., McCarthy, M.A., McDonnell, M. J., 2006. Local extinction of grassland plants: the landscape matrix is more important than patch attributes. Ecology 87, 3000-3006.

Wu, J., Boggess, W., 1999. The Optimal Allocation of Conservation Funds. Journal of Environmental Economics and Management 38, 302-321.

**Figures**

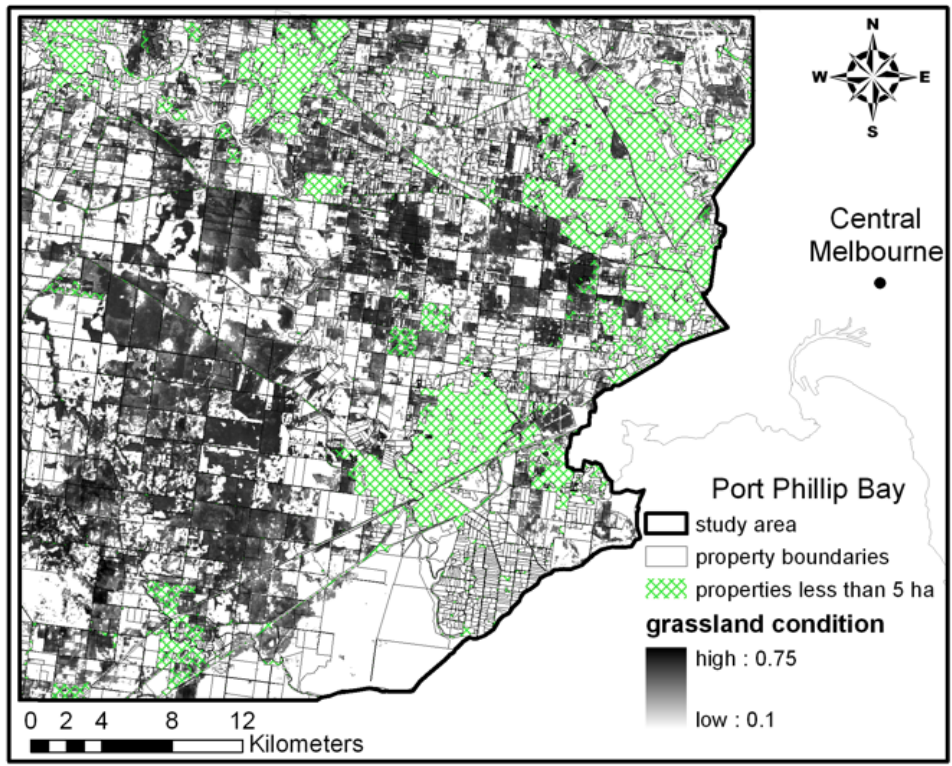

Fig. 1: The study area to the west of Melbourne, Australia. The black lines show land parcels in the study area and grassland condition is shown with graduated grey scale where a darker colour represents higher condition. The cross-hatched areas show the locations within which parcels less than 5 hectares occur. A high-resolution version of this figure is available in the Supplementary information (Fig. S6).

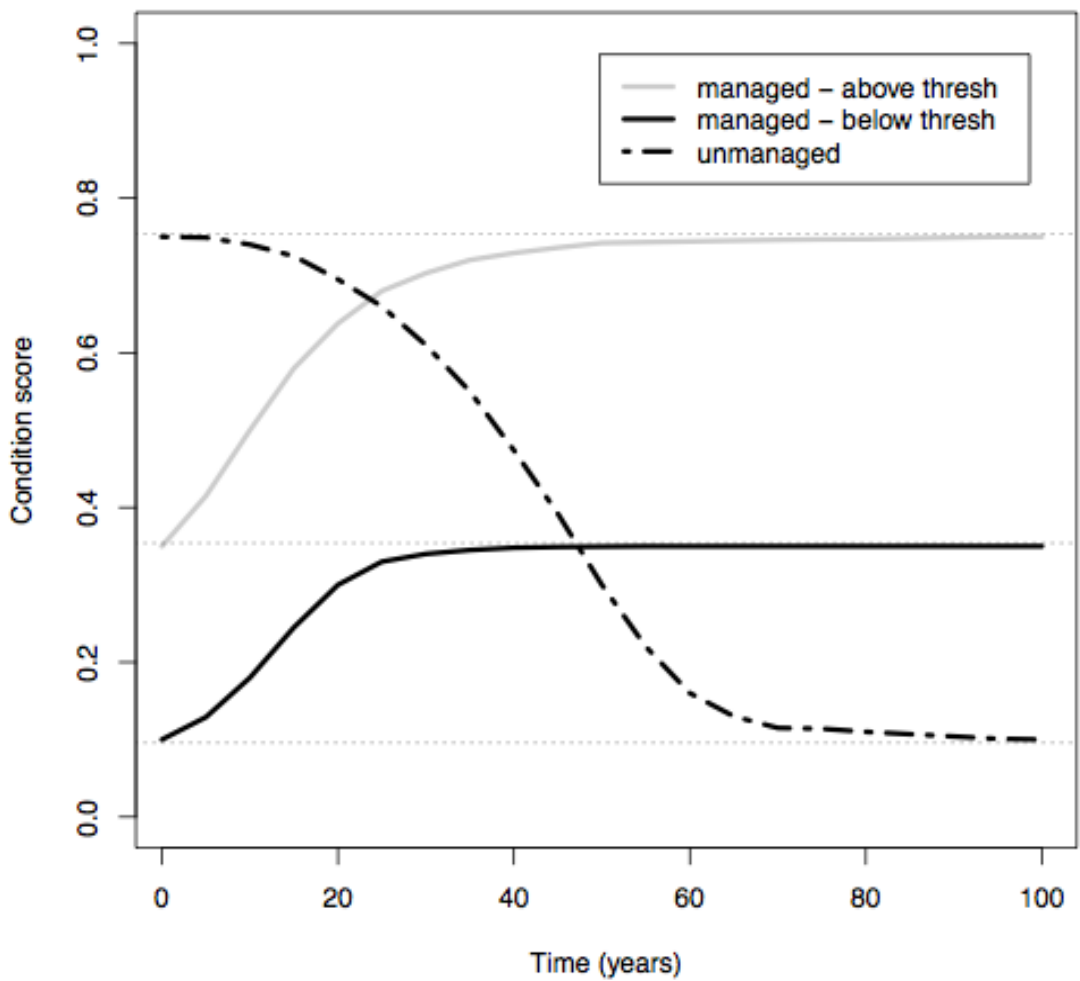

Fig. 2: Curves used to parameterise the grassland condition model and predict condition change over time for managed (solid lines) and unmanaged (dashed line) native grassland. Condition is measured in rescaled habitat hectares (see text). The solid black line depicts the condition change for vegetation already below the 0.35 threshold value. The solid grey line depicts the condition change for vegetation whose condition is above the 0.35 threshold value.

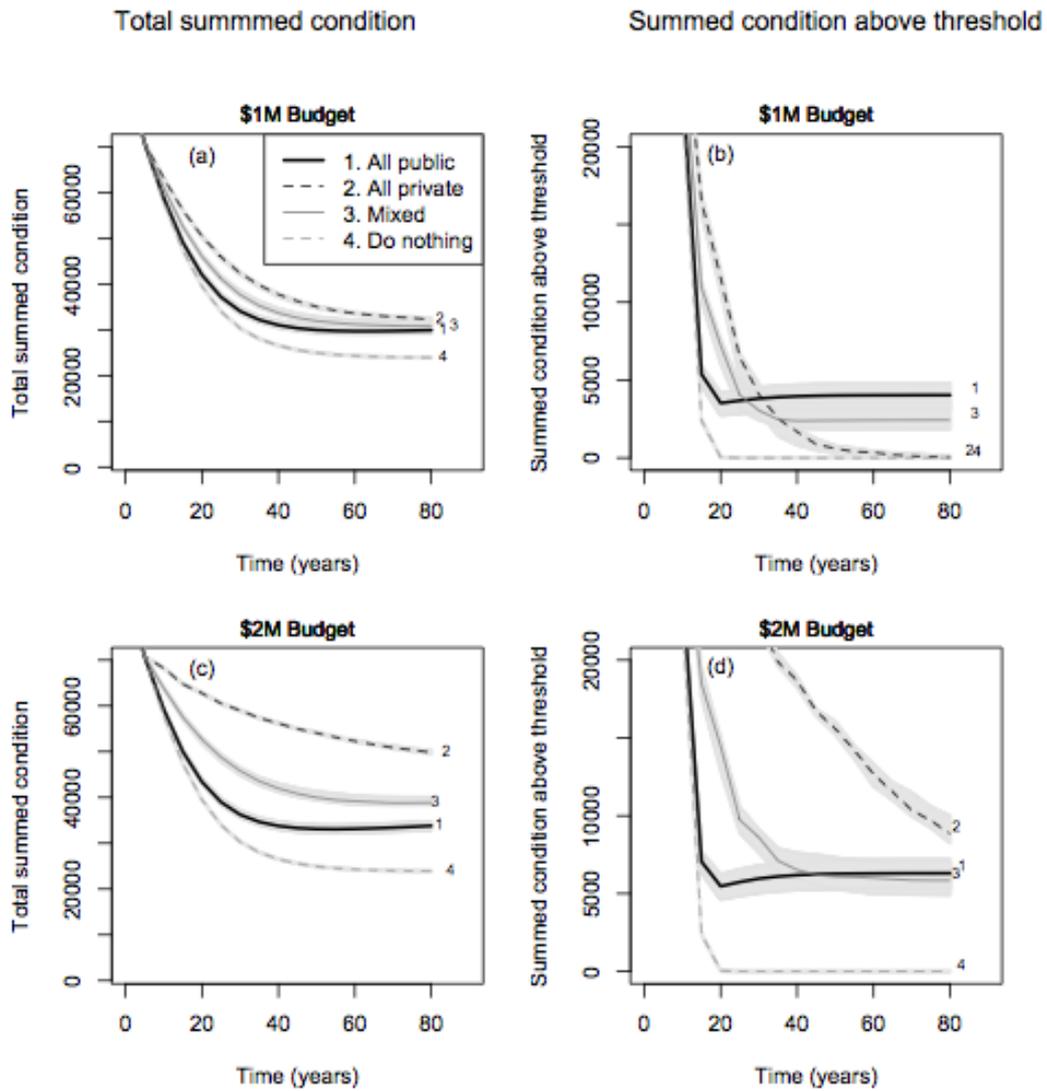

Fig. 3: Comparison of the four conservation policies with budgets of $1M and $2M per time-step, no development and *low* management costs. Plots in the left column show the total summed condition (TSC) as a function of time, while the plots in the right column show the summed condition above the condition threshold (CAT) (see text). The grey bands show the performance range and the line in the middle of each band shows the median performance over 5 realisations.

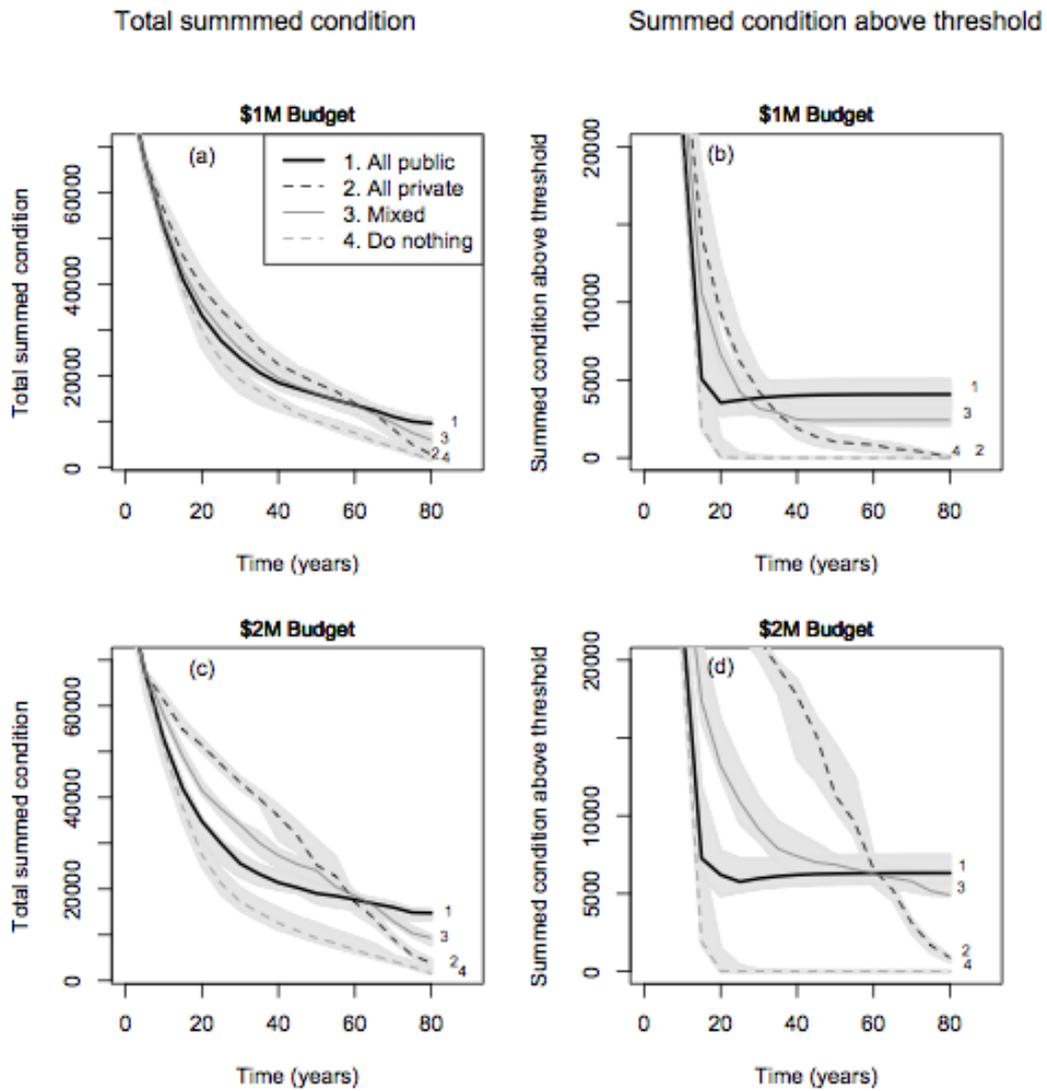

Fig. 4: Comparison of the four conservation policies with budgets of $1M and $2M per time-step with development occurring and *low* management costs. The arrangement of the plots is identical to Fig. 3.

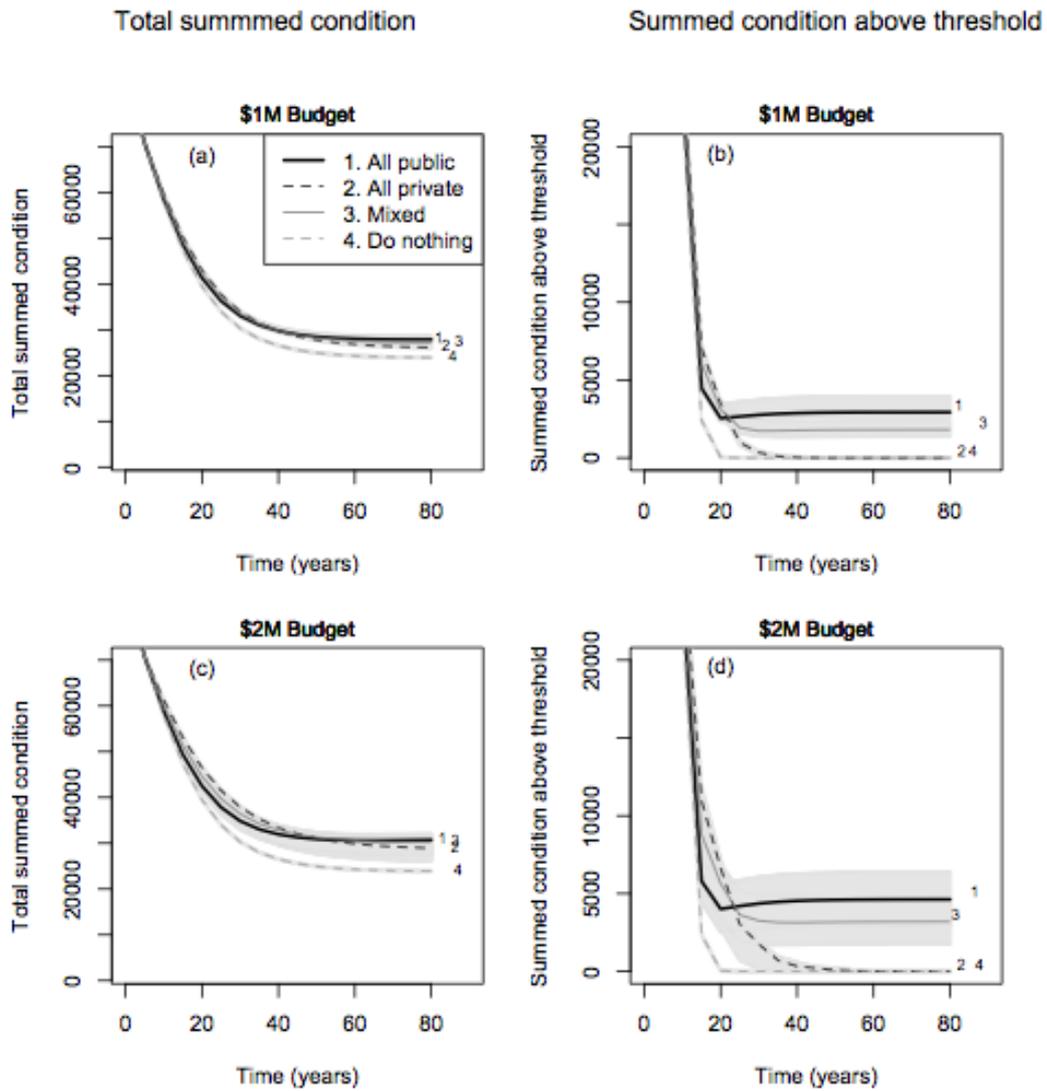

Fig. 5: Comparison of the four conservation policies with budgets of $1M and $2M per time-step using *high* management cost and no development. The arrangement of the plots is identical to Fig 3.

**Appendix A: Supplementary information**

*Background information on Western (Basalt) Plains Natural Temperate Grassland*

Native grasslands across temperate south eastern Australia have been modified structurally and have declined in species richness as a consequence of the interplay between a range of historic and ongoing processes including cultivation, intensive stock grazing, rabbit grazing, the addition of fertilisers and exotic plant introductions and invasions (McDougall and Kirkpatrick, 1993; Dorrough et al., 2004). Prior to European settlement these grasslands were regularly burnt by aboriginal people (Gott, 1994) and were subject to grazing by marsupials such as kangaroos. Contemporary native grassland remnants are in a range of conditions and contexts as a consequence of site-specific idiosyncrasies and their recent history (Lunt, 1997).

There have been measurable declines in the species diversity and condition of native grasslands that result from a range of factors. These include i) exotic plant invasion, particularly invasions mediated by perennial grasses (McLaren et al., 1998; Morgan 1998), ii) the absence of regular fire, grazing or drought which can result in acute interspecific competition exerted by the dominant tussock grass (Lunt and Morgan, 1999; Williams et al., 2006), iii) high grazing intensities and phosphate fertiliser application (Dorrough and Scroggie 2008, McIntyre and Lavorel 1994).

*Determining land parcel purchase cost*

Sale price data for the study was taken from a confidentialised extract of unit-record property sale valuations from the 2008 Victoria Valuer General Statewide Valuations Dataset. Property sales prices were selected if they had a rural land use categorisation, were in a municipality that overlapped or adjoined the study area and had been sold after 1990. All prices had been inflated from their nominal value to 2008 dollars.

As larger parcels tend to have a lower price per hectare, we split the sale prices into two samples based on whether the parcel had an area less than or greater than 150 hectares. Due to the small number of large parcels sold, there was not enough data for a finer scale binning based on parcel area. A lognormal function was then fitted to the distributions of prices per hectare to obtain two continuous functions from which to sample price data (Fig. S1).

At the start of each simulation every parcel was assigned a purchase price in the following manner. A price per hectare was determined by sampling from the appropriate lognormal distribution (Fig. S2) depending on whether the parcel had an area less than or greater than 150 hectares. The total price of the parcel was then determined by multiplying this price by the area of the parcel.

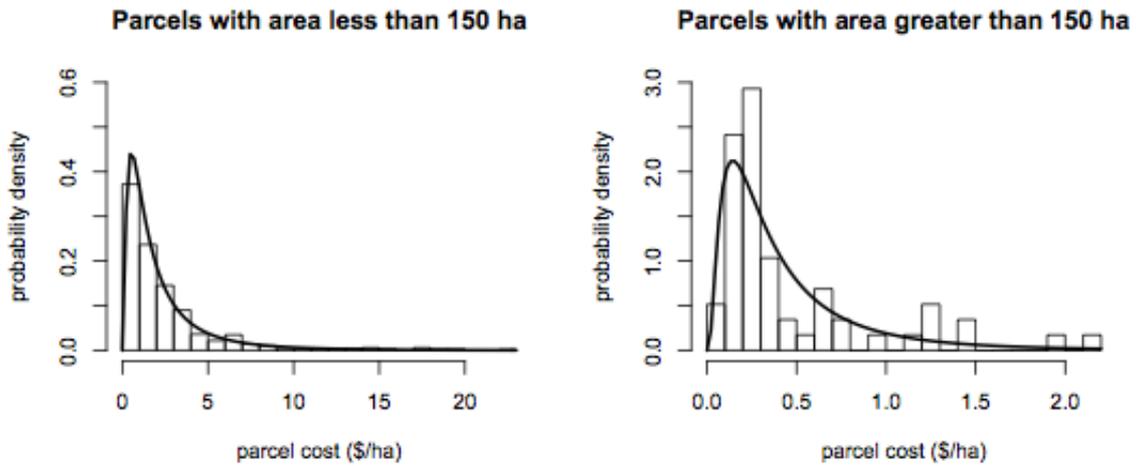

**Fig. S1.** The results from fitting a lognormal distribution to the property price data. Two separate fits were made for properties less than 150 ha (n = 688) and greater than 150 ha (n = 58).

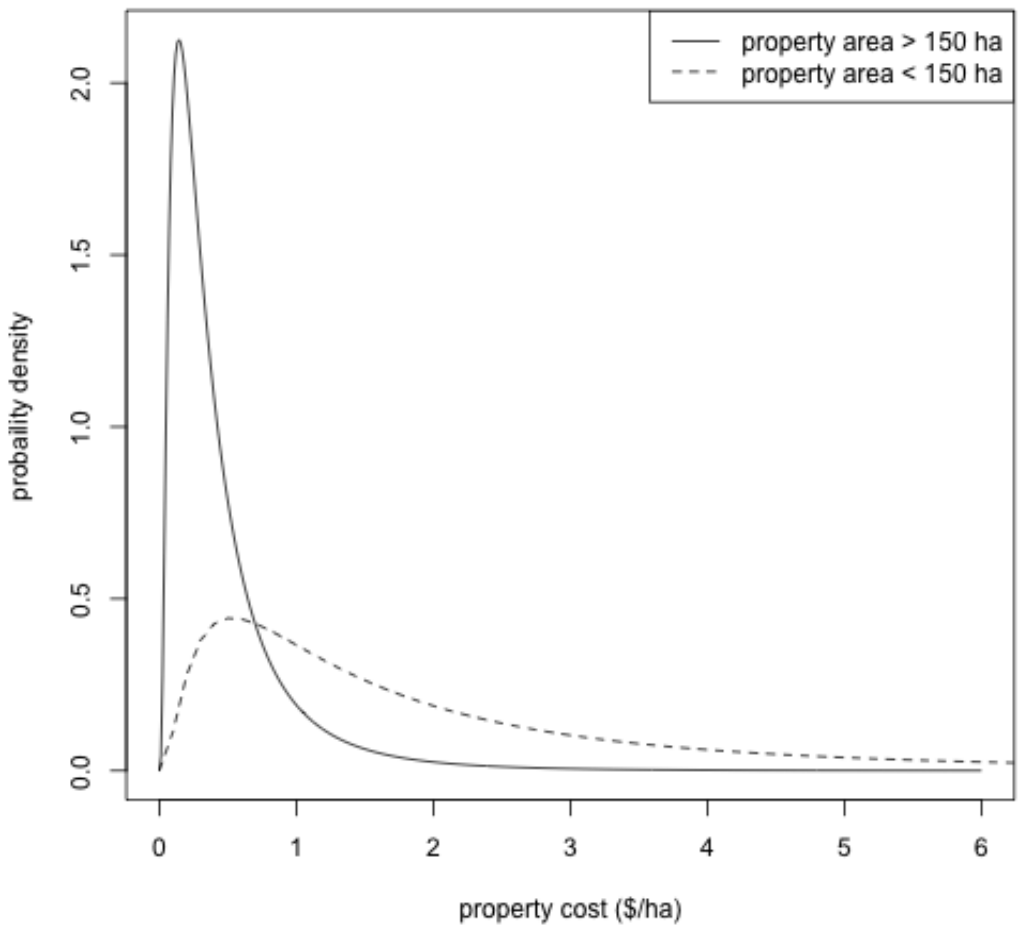

**Fig. S2.** The lognormal distributions used in determining parcel costs per hectare. The solid and dashed distributions were used for parcels with areas greater or less than 150 hectares, respectively. The total cost of a parcel was then obtained by multiplying the parcel cost per hectare by the area of the parcel.

*Further details on the grassland content of parcels*

Parcels with an area less than 5 hectares were excluded from the study, as these parcels were primarily residential and contain a small amount of lower quality grassland. These smaller parcels also tend to be aggregated into clusters that occur around the edges of the areas of higher quality grassland (Fig. S6). Table S1 gives further information regarding the grassland content of parcels above and below five hectares in size.

The parcels greater than 5 hectares cover 68% of the study area and contain 93% of the grassland in the study area. The grassland in parcels with an area less than five hectares tends to be in a lower condition with a median condition score of 0.36 versus a median of 0.45 for parcels greater than 5 hectares. Another aggregate measure of the condition is the proportion of grassland above the condition threshold. This measures the amount of the grassland that has the potential to be restored to a high condition (see Section 2.5 in the manuscript). Using this measure, 54% of the grassland in parcels less than five hectares is above the condition threshold compared to 71% for the parcels greater than five hectares. This means that only 5% of the higher-quality grassland in the study area is to be found on the parcels below 5 hectares in area.

**Table S1**. Further details on the grassland content of the parcels above and below 5 hectares in size.

|  | All parcels > 5 ha | All parcels ≤5 ha |
|---|---|---|
| Proportion of study area | 0.684 | 0.316 |
| Proportion of all grassland in study area | 0.933 | 0.067 |
| Median condition of grassland | 0.450 | 0.364 |
| Proportion of grassland above the condition threshold | 0.710 | 0.535 |
| Proportion of all grassland in study are above condition threshold | 0.949 | 0.051 |

Note: condition is measured in rescaled habitat hectares (see Section 2.2 in the manuscript).

*Results including $1.5M budget*

Figs. S3 – S5 show the same plots as Figs. 3 – 5 in the manuscript with additional results for the intermediate budget of $1.5M. As expected, the $1.5M budget results are consistent with a transition between the results for the $1M and $2M budgets.

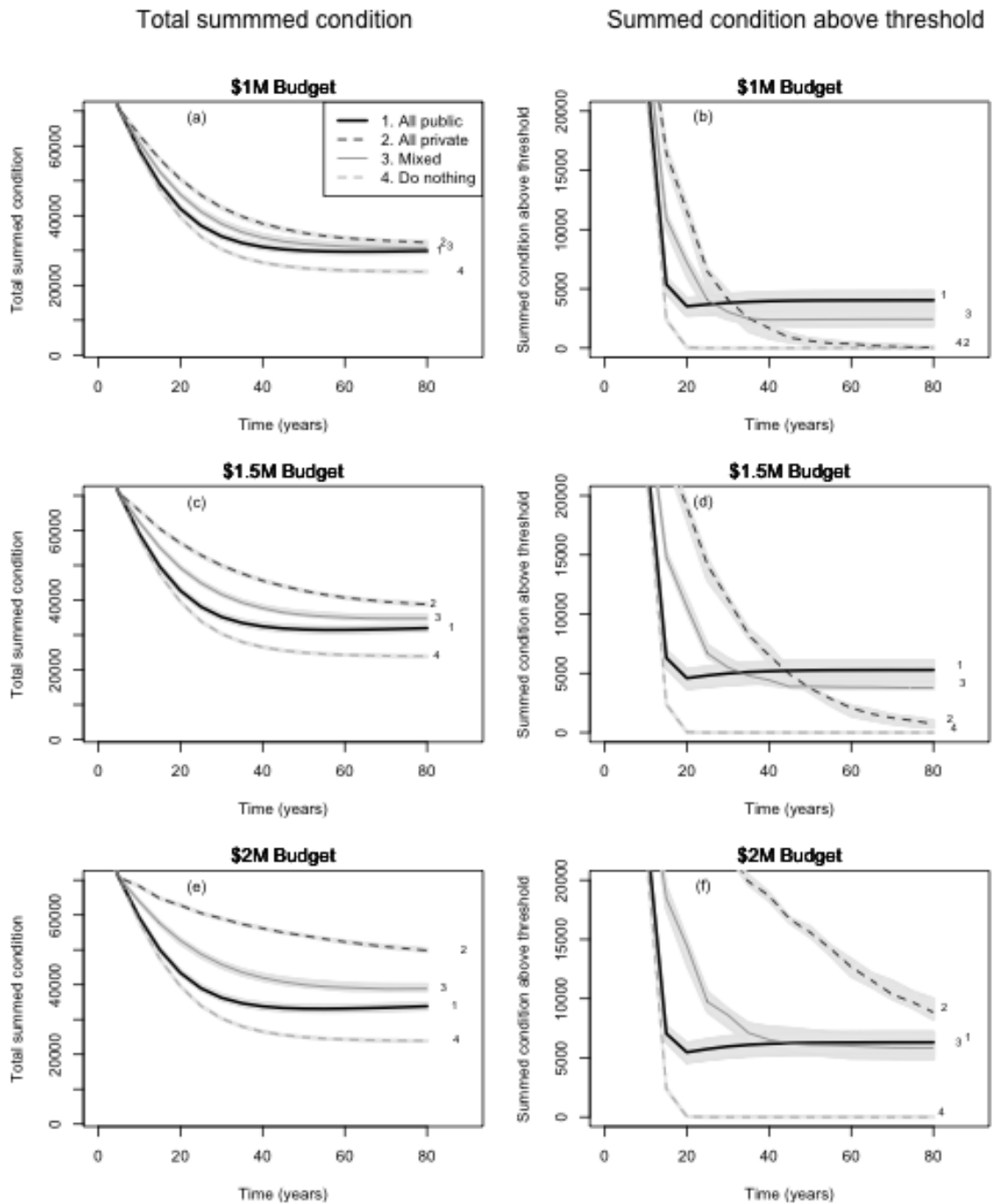

Fig. S3: Comparison of the four conservation policies with budgets of $1M, $1.5M and $2M per time step with no development and *low* management costs. Plots in the left column show the total summed condition as a function of time, while the plots in the right column show the summed condition above the condition threshold (see text). On each graph, all policies were run 5 times to depict the effects of model stochasticity. The grey bands show the performance range and the line in the middle of each band shows the median value over the 5 realisations.

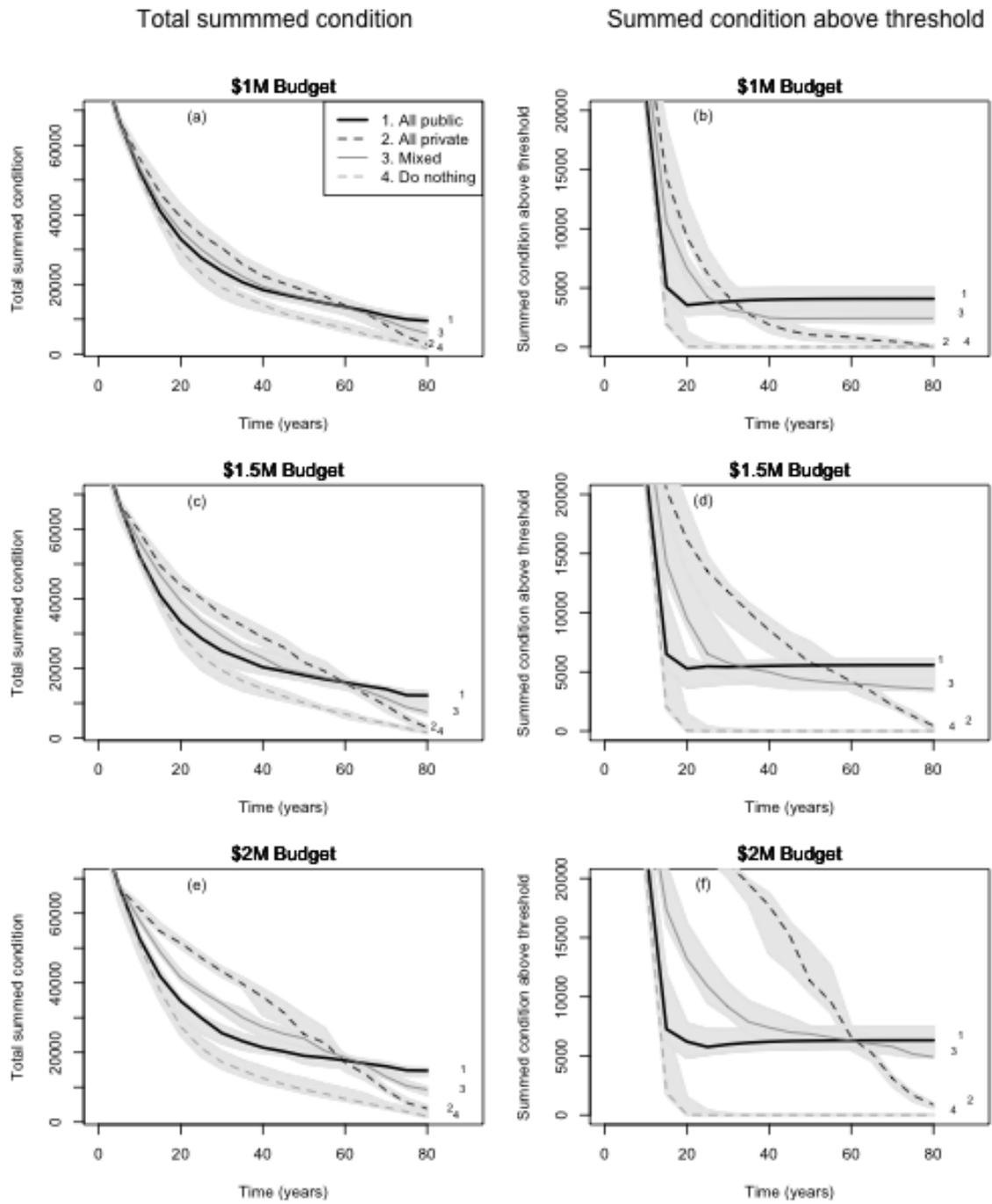

Fig. S4: Comparison of the four conservation policies with budgets of $1M, $1.5M and $2M per time step with development occurring and *low* management costs. The arrangement of the plots is identical to Fig. S3.

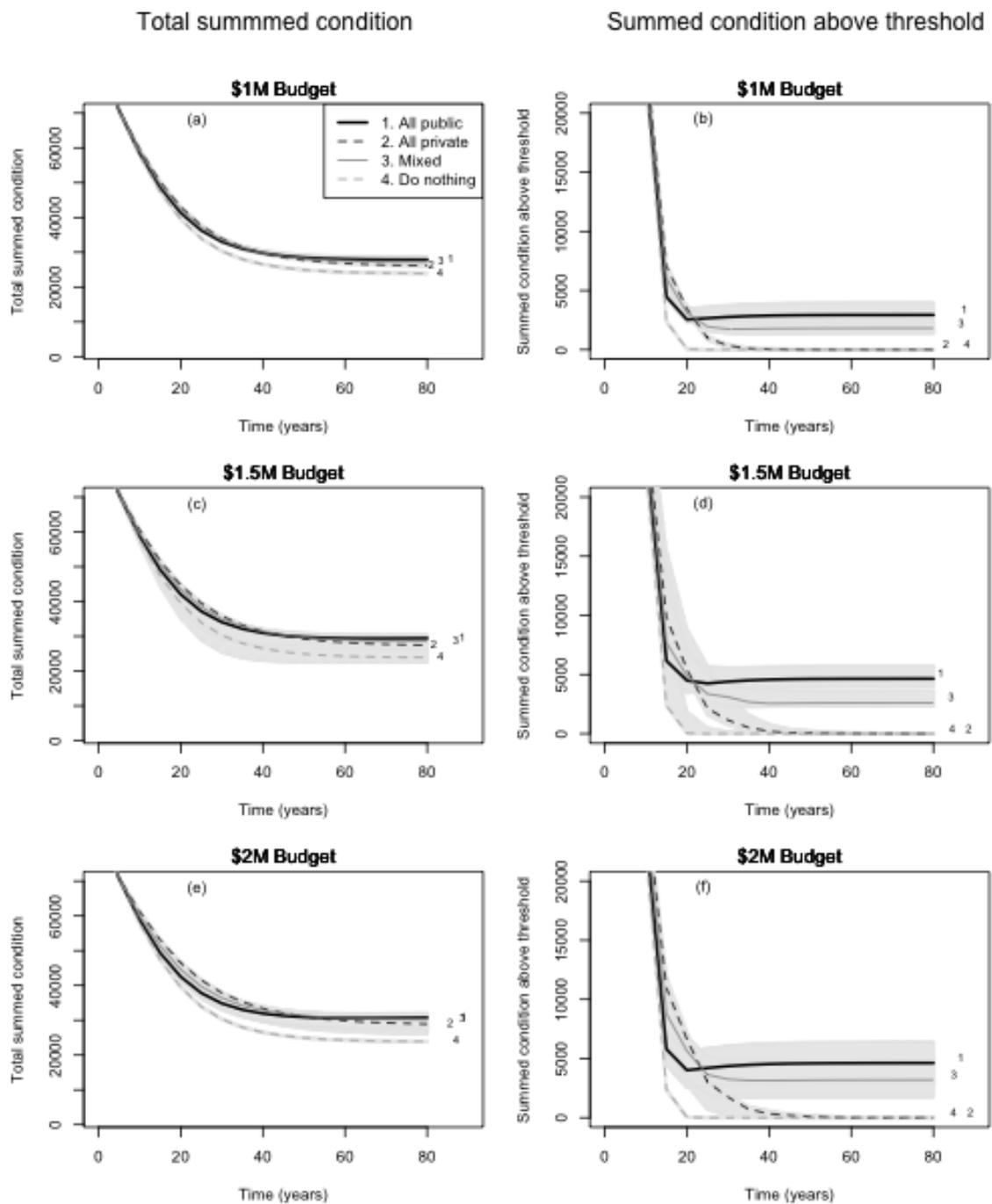

Fig. S5: Comparison of the four conservation policies with budgets of $1M, $1.5M and $2M per time step using the *high* management cost and no development. The arrangement of the plots is identical to Fig S3.

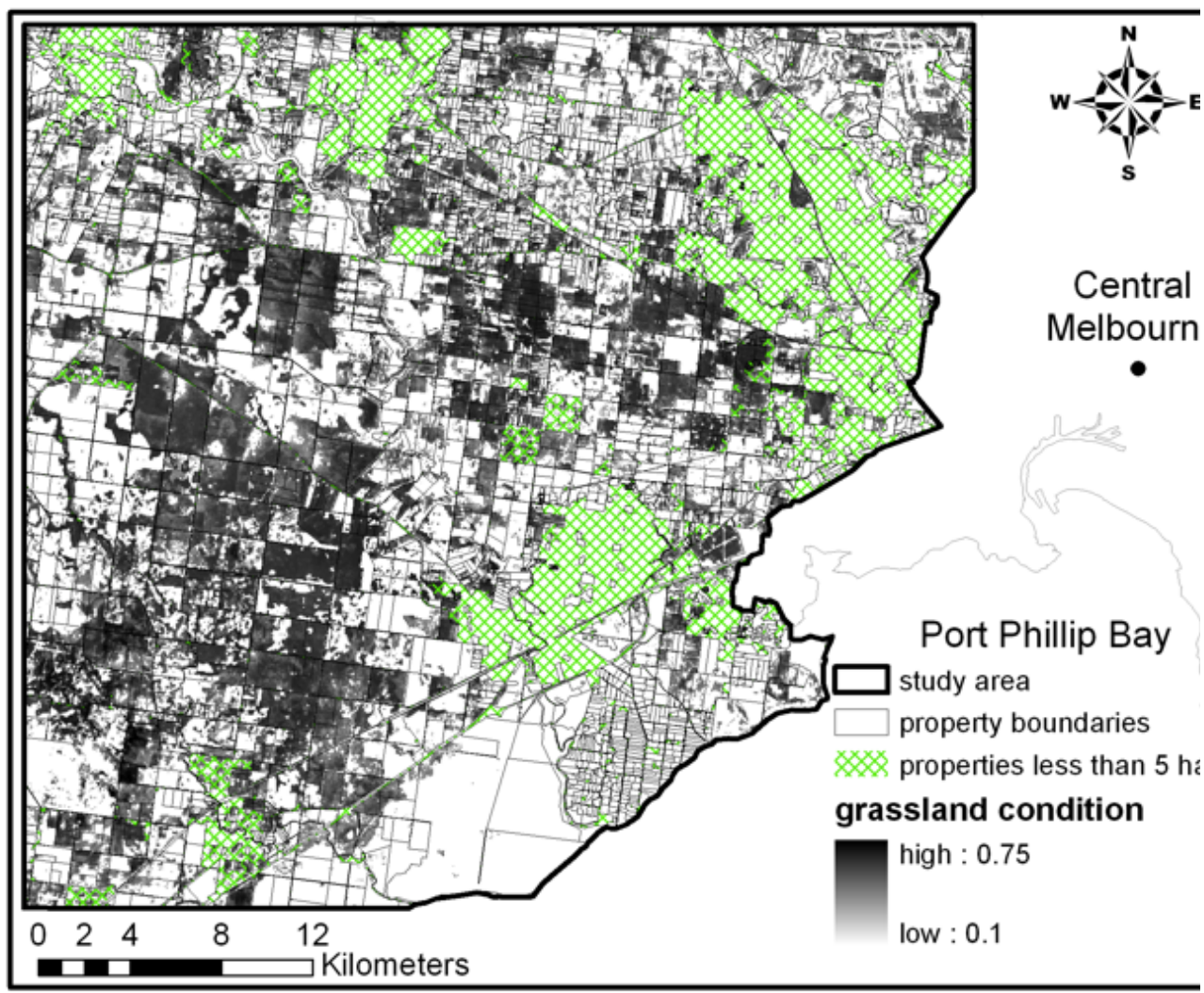

Fig. S6: The study area to the west of Melbourne, Australia. This is a higher resolution version of Fig. 1 in the manuscript. The black lines show land parcels in the study area and grassland condition is shown with graduated grey scale where a darker colour represents higher condition. The cross-hatched areas show the locations within which parcels less than 5 hectares occur.